\documentstyle[psfig,prb,aps]{revtex}
\draft

\title{Broken  Generalized Kohn Theorem in Harmonic Dot Lattices due to 
Coulomb Interaction between the Dots: 
Exact solution of the Schr\"odinger equation in the dipole approximation}
\author{by M. Taut\\Institute for Solid State and Materials Research 
Dresden\\ POB 270016,
01171 Dresden, Germany\\
email: m.taut@ifw-dresden.de\\}

\begin{document}
\maketitle

\begin{abstract}
The Generalized Kohn Theorem in arrays of parabolic quantum dots with
Coulomb interaction between the dots  is violated, if
there are different dot species involved.
We solve the Schr\"odinger equation for  cubic lattices with 
two different dots per unit cell:
i) two different circular dots and ii) two elliptical
dots, which are rotated by $90^o$ relative to each other.
The interaction between the dots is considered in dipole approximation and
long-- wavelength excitation spectra including FIR intensities are calculated.
The energy spectrum of the first case can be expressed as a superposition of two
noninteracting dots with an effective confinement frequency, which 
includes the effect of dot interaction.
Only in the second case a splitting of degenerate
absorption lines  and an anticrossing occurs,
which is a qualitative indication  for interdot interaction.
If the interaction becomes very strong and if all lattice sites (not necessarily
confinement potentials) are equivalent, then the contribution of the dot
interaction outweigh possible differences in the confinement potentials
and the Kohn Theorem gradually reentries, in the sense that one pair 
of excitation modes (pseudo Kohn modes) becomes 
independent of the interaction strength.

\end{abstract}

\section{Introduction}
The Generalized Kohn Theorem \cite{Kohn} (GKTh) plays a 
crucial role in quantum dot physics
with far reaching consequences.
It considers interacting electron systems in a harmonic confinement
and a constant magnetic field, and it states that excitations by 
long wavelength radiation 
are not effected by the electron electron (e e)
interaction. This statement applies to arrays of {\em identical}
 harmonic dot confinements
(with e e interaction between the dots) as well (see Appendix 
of \onlinecite{Taut-dot-lattice}).
This does not mean that all excitations are independent of e e interaction,
but only the optically active ones (Kohn modes), 
and it does not mean that e e interaction
is not important for the other excitations. However, this fact prevents
the e e interaction from beeing seen and  investigated e.g. by far infrared 
(FIR) spectroscopy. The FIR absorption spectrum of the whole system agrees 
exactly with the spectrum of a single particle.
The GKTh does not hold for arrays of {\em different} dot confinements, 
e.g. periodic
dot lattices with two different harmonic dot confinements per 
unit cell \cite{Taut-dot-lattice}. Then, all collective modes
are excited by FIR radiation  and effected by e e interaction, 
or in other words, there is no Kohn mode.
The calculation and investigation of absorption frequencies and probabilities 
in the latter case is the subject of this work.\\
In order to obtain a visual picture, let us first consider 
a classical model for the Kohn mode for vanishing magnetic field.
(This preliminary consideration will be replaced by a rigorous
 quantum mechanical treatment
in the following.)
 Classically, the charge distributions of all dots
oscillate rigidly in-- phase with the bare confinement frequency,
and the e e interaction contributes only a constant term to 
the total energy (independent of elongation).
If we have more than one {\em identical} dots per unit  cell, 
there are additional collective modes,
in which the individual dots oscillate out of phase, and which are affected by
dot interaction, but which are not optically active.
Consequently, the dot interaction is not observable 
with FIR spectroscopy in arrangements of {\em identical} dots.
One way to trick Kohn's theorem is to include {\em different} dot species.
Then, there is no coherent oscillation mode for all dots, which does not
change the e e interaction energy 
of the system in elongation, because there is
no common bare confinement frequency. As a consequence, 
all collective modes (two modes per dot in the unit cell) are effected  by 
dot interaction and excited by FIR radiation with a finite probability.
In other words, the Generalized Kohn Theorem  for dot arrays is broken.\\
Other systems, where Kohn's Theorem does not hold,
 comprise: i) anharmonic confinements \cite{Heitmann,Pfannkuche}
(circular dots with $r^4$ and higher order terms in the radial dependence or
cubic dots with terms of type $x^2 \;y^2$) , ii) 
hole dots with different effective masses \cite{Roessler}.
One point of this paper is that the GKTh can be broken despite an exactly
harmonic Hamiltonian.
A further possibility to observe the e e interaction in the 
excitations is to consider finite wave length 
\cite{Taut-dot-lattice,Heitmann}.
\\

\section{Magnetophonon Hamiltonian}
The first part of the calculation of the eigenstates of the Hamiltonian
 follows closely the procedure described in 
Ref.\onlinecite{Taut-dot-lattice}. We only have to consider that {\em now
 the confinement potentials and electron numbers can be different} 
in different dots.
After introducing center-- of -- mass (c.m.) and relative coordinates
in each dot and applying the dipole approximation for the Coulomb
interaction between the dots, we observe that the Hamiltonian of 
all c.m. coordinates is decoupled from individual dot Hamiltonians in the 
relative coordinates. That's why all excitations can be classified into
i) collective (c.m.) excitations, and  ii) intra-dot excitations. The latter are not considered here because they are not optically active.
The  Hamiltonian in the c.m. coordinates ${\bf R}_{n,\alpha}$ reads
in atomic units $\hbar=m=e=1$  
(see also Sect. IV A in Ref.\onlinecite{Taut-dot-lattice})
\begin{eqnarray}
H_{c.m.}&=&
 \sum_{n,\alpha}
\frac{1}{2 m^*}
\left[
\frac{{\bf P}_{n,\alpha}}{\sqrt{N_\alpha}}+
{\sqrt{N_\alpha} \over c} {\bf A}\left({\bf U}_{n,\alpha} \right)
\right]^2  \nonumber \\
 && + \frac{1}{2} \sum_{n,\alpha\atop n',\alpha'}\; 
\sqrt{N_\alpha N_{\alpha'}} \;\;\;{\bf U}_{n,\alpha} \cdot
{\bf C}_{n,\alpha;\, n',\alpha'}
\cdot {\bf U}_{n',\alpha'}
\label{H-cm-latt}
\end{eqnarray}
where ${\bf U}_{n,\alpha}={\bf R}_{n,\alpha}-{\bf R}_{n,\alpha}^{(0)}$
is the elongation of the c.m.
at lattice site $(n,\alpha)$ and  
${\bf P}_{n,\alpha}=-i\;{\bf \nabla} _{{\bf U}_{n,\alpha}} $ is
 the corresponding
canonical momentum operator. $n$ runs over the unit cells and $\alpha$
over the dot species within a cell. $N_\alpha$ is the number of electrons 
in dot $\alpha$,
 and $m^*$ the effective mass.
It is clear already from inspection of (\ref{H-cm-latt}) that 
the eigenvalues of $H_{c.m.}$ do not depend on the explicitly shown
electron numbers $N_\alpha$, because the factors $\sqrt{N_\alpha}$
can be considered just as a rescaling factor 
of the coordinates ${\bf U}_{n,\alpha}$.
However, the eigenfunctions (and quantities derived from them) do
depend on the explicit $N_\alpha$.
The force constant tensor reads
\begin{eqnarray}
C_{n,\alpha;\, n,\alpha}&=&
 {\bf \Omega}_{\alpha} + \epsilon^{-1} N_\alpha \sum_{n',\alpha'(\ne n,\alpha)}
{\bf T}\left( {\bf R}_{n,\alpha}^{(0)}-
{\bf R}_{n',\alpha'}^{(0)} \right)   \label{C1-latt}\\
C_{n,\alpha;\, n',\alpha'}&=&
-\epsilon^{-1} \sqrt{N_\alpha N_{\alpha'}} \;{\bf T}\left( {\bf R}_{n,\alpha}^{(0)}-
{\bf R}_{n',\alpha'}^{(0)} \right)~~~
\mbox{for} ~~~(n,\alpha) \ne (n',\alpha')
\label{C2-latt}
\end{eqnarray}
where $\epsilon^{-1}$ is the inverse background
dielectric constant
and  ${\bf \Omega}_\alpha$ the bare confinement tensor, 
which produces a harmonic confinement.
The dipole tensor is defined as
$
{\bf T}(\mbox{\boldmath $a$})=\frac{1}{a^5}\;\left [ 3 \; \mbox{\boldmath $a$}
\circ \mbox{\boldmath $a$} - a^2\;
{\bf I} \right ]
$
where ($\circ$) denotes  the dyad product  and  $\bf I$ the unit tensor.
Observe that $\bf C$ depends on $N_\alpha$ implicitely which effects
the energy eigenvalues.

A unitary transformation to collective magnetophonon coordinates 
\begin{eqnarray}
{\bf U}_{n,\alpha}&=&\frac{1}{\sqrt{N_c}} \sum_{\bf q}^{BZ}
e^{-i{\bf q}\cdot R_n^{(0)}}\;{\bf U}_{{\bf q},\alpha}\\
{\bf P}_{n,\alpha}&=&\frac{1}{\sqrt{N_c}} \sum_{\bf q}^{BZ}
e^{+i{\bf q}\cdot R_n^{(0)}}\;{\bf P}_{{\bf q},\alpha}
\label{phonon-trafo}
\end{eqnarray}
where $N_c$ is the number of unit cells, leaves us with
a sum on $N_c$ decoupled subsystems 
$H_{c.m.}= \sum_{\bf q}\;H_{\bf q}$
\begin{eqnarray}
H_{\bf q}&=&   \sum_\alpha\; \frac{1}{2 m^*}
\left[
\frac{{\bf P}_{{\bf q},\alpha}}{\sqrt{N_\alpha}}
+\frac{\sqrt{N_\alpha}}{c}{\bf A}({\bf U}_{{\bf q},\alpha}^*)
\right]^\dagger \cdot
\left[
\frac{{\bf P}_{{\bf q},\alpha}}{\sqrt{N_\alpha}}
+\frac{\sqrt{N_\alpha}}{c}{\bf A}({\bf U}_{{\bf q},\alpha}^*)
\right] \nonumber\\
& &+ \frac{1}{2} \sum_{\alpha,\alpha'}
\sqrt{N_\alpha N_{\alpha '}}\;\;\;
{\bf U}_{{\bf q},\alpha}^* \cdot
{\bf C}_{{\bf q};\alpha,\alpha'}
\cdot {\bf U}_{{\bf q},\alpha'}  
\label{H-phonon}
\end{eqnarray}
which includes the dynamical matrix 
\begin{equation}
{\bf C}_{{\bf q};\alpha,\alpha'}=\sum_n\;
e^{i{\bf q}\cdot {\bf R}_n^{(0)}}\;
{\bf C}_{\alpha,\alpha'}\left({\bf R}_n^{(0)}\right)
~;~~~~{\bf C}_{\alpha,\alpha'}\left({\bf R}_n^{(0)}\right)
= {\bf C}_{n,\alpha;\,0,\alpha'}
\label{dyn-mat-full}
\end{equation}
With (\ref{C1-latt}) and (\ref{C2-latt}), we obtain
\begin{eqnarray}
C_{{\bf q};\alpha,\alpha}&=&
 {\bf \Omega}_{\alpha} + \epsilon^{-1} N_\alpha \sum_{\alpha'(\ne \alpha)}
{\bf T}\left( \mbox{\boldmath $a$}_{\alpha}-
\mbox{\boldmath $a$}_{\alpha'} \right) \\
&+& \sum_{n \ne  0}
\left[
\sum_{\alpha'} {\bf T}\left( {\bf R}_n^{(0)}+ \mbox{\boldmath $a$}_{\alpha}-
\mbox{\boldmath $a$}_{\alpha'} \right)
-e^{i{\bf q}\cdot {\bf R}_n^{(0)}} \; {\bf T}\left( {\bf R}_n^{(0)}\right)
\right]\\
C_{{\bf q};\alpha,\alpha'}&=&
-\epsilon^{-1} \sqrt{N_\alpha N_{\alpha'}}\; \sum_n e^{i{\bf q}\cdot {\bf R}_n^{(0)}}
{\bf T}\left( {\bf R}_n^{(0)}+ \mbox{\boldmath $a$}_{\alpha}-
\mbox{\boldmath $a$}_{\alpha'} \right) \;\;\;  \mbox{for} 
\;\;\;\alpha \ne \alpha'
\end{eqnarray}
where ${\bf R}_{n,\alpha}^{(0)}={\bf R}_n^{(0)}+\mbox{\boldmath $a$}_{\alpha}$
and $n\ne 0$ under the sum means that 
the term ${\bf R}_n^{(0)}=0$ is excluded.\\ 

\begin{figure}[h]
\vspace{-2cm}
\begin{center}
{\psfig{figure=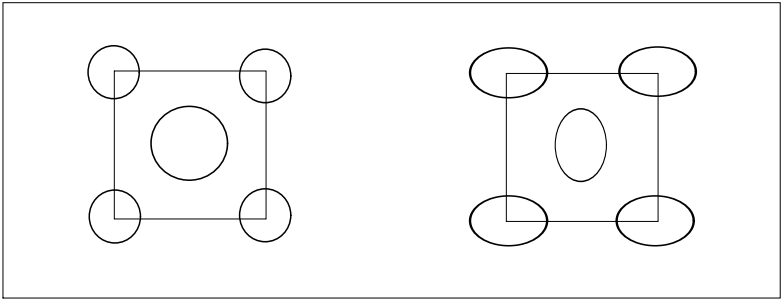,angle=-90,width=12.cm,bbllx=15pt,bblly=45pt,bburx=580pt,bbury=750pt}}
\caption[ ]{
Minimum unit cells for the two dot architectures considered in this paper 
with two
different circular dots (left) and two identical, but rotated, 
ellipsoidal dots (right).
}
\label{Fig1}
\end{center}
\end{figure}

Now, we  focus our attention to long-- wavelength modes 
(the index ${\bf q}=0$ is dropped henceforth)
 and consider a simple cubic lattice, alternatively occupied by two different 
dot species.
The minimum unit cell is face centered cubic (see Fig.1) 
with lattice constant $a$. After performing the lattice sum involved in 
(\ref{dyn-mat-full}) numerically, we obtain the four $2\times2$ dynamical matrices
\begin{eqnarray}
{\bf C}_{11}&=&{\bf \Omega}_1+d\; p_1\; {\bf I}\\
{\bf C}_{22}&=&{\bf \Omega}_2+d\; p_2\; {\bf I}\\
{\bf C}_{12}&=&{\bf C}_{21}=-d\; p_{12}\; {\bf I}
\end{eqnarray}
with the interaction parameters 
\begin{eqnarray}
p_i&=&2 N_i \epsilon^{-1} / (n.n.distance)^3=
4 \sqrt{2} N_i \epsilon^{-1} / a^3, \;\;\;\;\;(i=1,2) \label{pi}\\ 
p_{12}&=&2\sqrt{N_1 N_2} \epsilon^{-1} / (n.n.distance)^3=
4 \sqrt{2} \sqrt{N_1 N_2} \epsilon^{-1} /a^3 \label{p12}
\end{eqnarray} 
and $d=1.460$. From the preceding definitions it 
follows that $p_{12}= \sqrt{p_1 p_2}$. \\

\section{Eigenstates}
Now we are going to find eigenvalues and eigenfunction of
(\ref{H-phonon}).
For avoiding divergences for $B=0$, we add an isotropic 
oscillator potential $\frac{1}{2}\; \sum_\alpha \omega_0^2 {\bf U}_\alpha^2$
to the kinetic energy in (\ref{H-phonon}) and subtract it from the 
interaction term. $\omega_0$ is in principle 
arbitrary, but we chose the mean value
of the bare confinement frequencies included in $\Omega_1$ and $\Omega_2$.
Now we replace the coordinates in (\ref{H-phonon}) (for ${\bf q}=0$)
by Boson ladder operators. This is analogous to the usual text book
transformation 
(see e.g. Ref.\onlinecite{Hawrylak-Buch} Sect. 3.3) 
apart from the factors $\sqrt{N_\alpha}$. It is obvious that 
this modification can be taken into account by 
introducing scaled coordinates ${\bf U}_\alpha \rightarrow
\tilde {\bf U}_\alpha =\sqrt{N_\alpha}\;{\bf U}_\alpha$
 (what implies ${\bf P}_\alpha \rightarrow
\tilde {\bf P}_\alpha ={\bf P}_\alpha /\sqrt{N_\alpha}$).

\begin{eqnarray}
\sqrt{N_\alpha}\; U_{\alpha x} &=&
\frac{1}{2}\sqrt{\frac{2}{\tilde\omega_c^*}} \; 
\bigg( a_{\alpha 1}^{+}  +a_{\alpha 2}^{+}  +a_{\alpha 1} + a_{\alpha 2}
 \bigg)\\
\sqrt{N_\alpha} \; U_{\alpha y} &=
&- \frac{i}{2}\sqrt{\frac{2}{\tilde\omega_c^*}} \;
\bigg( a_{\alpha 1}^{+}  -a_{\alpha 2}^{+}  -a_{\alpha 1} + a_{\alpha 2} \bigg)
\end{eqnarray}
where  the first subscript $(\alpha=1,2)$ indicates the 
dot number and the second one the component.
The transformation of the  c.m. momentum operators is analogous.
\begin{eqnarray}
\frac{P_{\alpha x}}{\sqrt{N_\alpha}} &=&
 \frac{i}{2}\sqrt{\frac{\tilde\omega_c^*}{2}} \;
\bigg( a_{\alpha 1}^{+}  +a_{\alpha 2}^{+}  -a_{\alpha 1} - a_{\alpha 2}
 \bigg)\\
\frac{P_{\alpha y}}{\sqrt{N_\alpha}} &=&
 \frac{1}{2}\sqrt{\frac{\tilde\omega_c^*}{2}} \;
\bigg( a_{\alpha 1}^{+}  -a_{\alpha 2}^{+}  +a_{\alpha 1} - a_{\alpha 2} \bigg)
\end{eqnarray}
The cyclotron frequency is $\omega_c^*=B/m^* c$  and
 $\tilde \omega_c^*=\sqrt{\omega_c^{*2}+4 \omega_0^2}$.
Firstly, it is clear that the Hamiltonian in these ladder operators does not
show an explicit $N_\alpha$-- dependence anymore (apart from that
implicit in the dynamical matrix). 
This implies that the eigenvalues of the Hamiltonian
do not depend on those $N_\alpha$ explicitly seen in (\ref{H-phonon}).
Secondly, the commutators of the ladder operators are not influenced by
the $N_\alpha$--factors and agree with those of Bosons: 
$[a_{\alpha i},a^+_{\alpha i}]=1$ and all other commutators vanish.
(This is because the
commutators of the $\tilde {\bf U}_\alpha$ and 
$\tilde {\bf P}_{\alpha'}$ agree with
the commutators of the untilded quantities.)
Now, the total Hamiltonian can be written in matrix notation in the 
following compact form
\begin{equation}
H=\left[ \mbox{\boldmath $a$}^+  \mbox{\boldmath $a$}
 \right] \cdot {\bf H} \cdot 
\left[ \begin{array}{c}  \mbox{\boldmath $a$} \\
  \mbox{\boldmath $a$}^+ \end{array} \right]
\label{H-compact}
\end{equation}
where
\begin{equation}
\left[\mbox{\boldmath $a$}^+ \mbox{\boldmath $a$}\right]=
\left[a_{11}^+\; a_{12}^+\; a_{21}^+\; a_{22}^+ \;|\;
 a_{11}\; a_{12}\; a_{21}\; a_{22} \right]
\label{a-line}
\end{equation}
and $\left[ \begin{array}{c} \mbox{\boldmath $a$} \\
 \mbox{\boldmath $a$}^+ \end{array} \right]$
 is the transposed and Hermitian conjugate
of (\ref{a-line}). The $8\times 8$ Hamiltonian matrix is not unique,
 but can be cast
into  the following form
\begin{equation}
{\bf H}= \left[\begin{array}{cc} \alpha \;\; \beta\\  \beta^* \;\; \alpha^*
         \end{array} \right] \;\;\;\mbox{with} \;\;\;
\alpha^+ =\alpha \;\; ; \;\; \beta^T=\beta
\label{H-mat}
\end{equation}
consisting of the $4\times4$ matrices 
\begin{equation}
\alpha=\frac{1}{2} \left[ \begin{array}{cc} \omega \;0 \\ 0\;\omega
\end{array} \right]
+\frac{1}{4\; \tilde \omega_c^*}\; {\bf E}^+ \cdot 
\left[ \begin{array}{cc}  \tilde{\bf C}_{11} \; {\bf C}_{12}\\
      {\bf C}_{21} \; \tilde{\bf C}_{22} 
\end{array} \right] \cdot {\bf E} \\
\end{equation}
\begin{equation}
\beta= \frac{1}{4\; \tilde \omega_c^*}\; {\bf E}^+ \cdot
\left[ \begin{array}{cc}  \tilde{\bf C}_{11} \; {\bf C}_{12}\\
      {\bf C}_{21} \; \tilde{\bf C}_{22}
\end{array} \right] \cdot {\bf E} ^*
\end{equation}
with 
$
{\bf E}=\left[ \begin{array}{cc} \varepsilon \;\;\; 0\\ 0\;\;\; \varepsilon
\end{array} \right]
$
and the $2\times2$ matrices
\begin{equation}
\tilde{\bf C}_{kk}={\bf C}_{kk} -\frac{1}{2} \; 
\omega_0^2\; {\bf I}  \;\;\;;\;\;\;
\omega=\left[ \begin{array}{cc} \omega_+ \;\;\;\; 0 \\
                                 0\;\;\;\;\;  \omega_- 
\end{array} \right]
\;\;\; ; \;\;\;
\varepsilon=\left[ \begin{array}{cc} 1\;\;\;\;1\\ \;\;i\;\;\;-i
\end{array} \right]
\end{equation}
with
\begin{equation}
\omega_\pm=\sqrt{\omega_o^2+\left(\frac{\omega_c^*}{2}\right)^2}\pm
\left(\frac{\omega_c^*}{2}\right)
\end{equation}
Finding the eigenstates of the Boson Hamiltonian (\ref{H-compact}) is provided
by mathematical physics and described in Ref.\onlinecite{Tsallis} in full detail. 
The goal is to find a linear transformation
$
\left[ \begin{array}{c} \mbox{\boldmath $b$} \\
 \mbox{\boldmath $b$}^+ \end{array} \right]={\bf A} \cdot 
\left[ \begin{array}{c} \mbox{\boldmath $a$} \\
 \mbox{\boldmath $a$}^+ \end{array} \right]
$
which preserves Boson commutators and diagonalizes $H$.
We shall only summarize the recipe here.\\
The {\em eigenvalues} are given by
$
E_{n_1,n_2,n_3,n_4}=\sum_k^{(1...4)} \;  \bigg( n_k+ \frac{1}{2} \bigg)\;
\omega_k
$
with $n_k$ being non--negative integers and 
$\omega_k=2\; \gamma_k$ with $\gamma_k$
being the four {\em positive} eigenvalues of the matrix 
${\bf H}\cdot{\bf J}$. The $8\times 8$ matrix
$
{\bf J}=\left[ \begin{array}{cc} 
{\bf I} \;\;\; 0\\ \;\; 0\;\;\;-{\bf I}
\end{array} \right]
$
is made up of $4\times4$ unit matrices. All eigenvalues of ${\bf H}\cdot{\bf J}$
come in pairs $(\gamma_k,-\gamma_k)$.\\
The {\em eigenfunctions} of $H$ are constructed as usual for Bosons         
\begin{equation}                                            
|n_1,n_2,n_3,n_4>=\prod_k^{(1...4)} 
\frac{\bigg( b_k^+ \bigg)^{n_k}}{\sqrt{n_k!}} \;|0>
\label{eigenvectors}
\end{equation}                                              \\
The four eigenvectors belonging to the positive eigenvalues 
are written in the form
${\bf x}_k=\left[ \begin{array}{c} \mbox{\boldmath $u$}_k \\
 \mbox{\boldmath $v$}_k \end{array} \right]$.
The column vectors of 
${\bf A}^+$ are given by the  vectors ${\bf x}_k$, and by the vectors
$\hat {\bf x}_k=\left[ \begin{array}{c} \mbox{\boldmath $v$}_k^* \\
 \mbox{\boldmath $u$}_k^* \end{array} \right]$,
which are the eigenvectors belonging to $-\gamma_k$.
The eigenvectors have  to be
properly orthonormalized ${\bf x}_i^+\cdot {\bf J}\cdot{\bf x}_k= \delta_{i,k}$.
Without degeneracy, the orthogonality is guaranteed automatically.
The inverse of this particular transformation is obtained from
${\bf A}^{-1}={\bf J} \cdot {\bf A}^+ \cdot {\bf J}$
which shows that the linear transformation is not unitary (but unitary
in a non-- Euklidian metric).

\section{Oscillator strength}
Optical oscillator strength between the states $|n>=|n_1,n_2,n_3,n_4>$ and
$|n'>=|n'_1,n'_2,n'_3,n'_4>$ for polarization in $\eta=(x\; \mbox{or}\; y)$ 
direction are defined as
\begin{equation}
f_{n,n';\eta}=2\;m^* \; \omega_{n,n'}\; 
|<n|{\bf U}_{\eta;tot} |n'>|^2
\label{trans}
\end{equation}
where $\omega_{n,n'}$ is the corresponding excitation energy, 
and ${\bf U}_{\eta;tot}$ is the  $\eta$--component of the 
total c.m. of the electrons in a unit cell 
(apart from a constant term).  
In formulae, this means ${\bf U}_{tot}=\frac{N_1}{N_{tot}}{\bf U}_{1}
+\frac{N_2}{N_{tot}}{\bf U}_{2}$, where $N_{tot}=N_1+N_2$.
After expressing the vectors ${\bf U}$ by ladder operators $b_k, b_k^+$ 
and using (\ref{eigenvectors}), we obtain 
the usual selection rules, i.e., only one quantum with energy $\omega_k$
can be absorped or emitted, so that we obtain only four 
absorption lines.
The result for the oscillator strength for the four possible 
transitions $(k=1...4)$
and for $\eta$-- polarization reads
\begin{equation}
f_{k,\eta}=\frac{m^*\;\omega_k}{N_{tot}\;\tilde\omega_c^*} \;
|S_{k,\eta}|^2 \cdot \bigg\{ \begin{array}{c} (n_k+1)
\\ n_k \end{array}\;\;
\mbox{for}\;\;
\bigg\{ \begin{array}{c}  absorption
\\ emission \end{array}
\end{equation} 
where $n_k$ denotes the initial state, and
\begin{eqnarray}
S_{k,x}&=&\sum_i^{(1,2)} \sqrt{\frac{N_1}{N_{tot}}}
 \bigg( u_{ki}-v_{ki}\bigg) 
+\sum_i^{(3,4)} \sqrt{\frac{N_2}{N_{tot}}}
 \bigg( u_{ki}-v_{ki}\bigg)\\
S_{k,y}&=&\sum_i^{(1,2)} \sqrt{\frac{N_1}{N_{tot}}}
(-1)^{(i+1)} \bigg( u_{ki}+v_{ki}\bigg)
+\sum_i^{(3,4)} \sqrt{\frac{N_2}{N_{tot}}}
(-1)^{(i+1)} \bigg( u_{ki}+v_{ki}\bigg)
\end{eqnarray}
In the last definition, $u_{ki}$ and $v_{ki}$ for i=1...4 are the components of 
the vectors ${\bf u}_k$ and ${\bf v}_k$, respectively.
The oscillator strength defined in (\ref{trans}) fulfill the following 
exact $f$--sum rule
 $\sum_k f_{k,\eta}= \frac{1}{N_{tot}}$. It is worth pointing out
 that for equal electron numbers in either dot ($N_1=N_2=N$),
the oscillator strength depends explicitly on $N$ (contrary to
the optical excitation energies). In all figures presented below
the oscillator strength are for $N_1=N_2=N$.\\ 

\section{Results}
Now the two simplest cases are discussed in more detail: 
two different circles and two
identical, but rotated ellipses. The ratio of the 
two bare confinement frequencies
involved in either case is 1:1.5 which means, that the two confinemnet 
frequencies in units of the mean frequency $\omega_0$ are 1.2 1nd 0.8. 
In our figures, all frequencies (energies)
are given in units of the mean confinement frequency $\omega_0$ and
the interaction parameters $p$ in units of $\omega_0^2$. The magnetic field
is given in terms of the effective cyclotron frequency $\omega_c^*$
in units of $\omega_0$ (upper scale)
and in $Tesla$ (lower scale).
The conversion between both scales is provided by
\begin{equation}
\omega_c^*[\omega_0]=\frac{0.9134\cdot10^{-2}}{m^*\;\omega_0[a.u.^*]}B[Tesla]
\end{equation}
In our figures we used $\omega_0=0.2 \;a.u.^*= 2.53\; meV$
and $m^*$ of GaAs for this conversion. (We want to stress
that this parameter choice effects only  the magnetic field scale
 and not  the curves.)
The definitions of the interaction parameters (\ref{pi})
for GaAs in more convenient units reads
\begin{equation}
p_i[\omega_0^2]=\frac{2.26 \cdot10^7 \; N_i}{\big(n.n.distance[\AA\big])^3\;
\big(\omega_0[meV]\big)^2}
\label{p-GaAs}
\end{equation}
(For a more detailed discussion of order-- of -- magnitude  estimates see
Ref. \onlinecite{Taut-dot-lattice}.)\\

For two different
{\em circular dots} with bare confinement frequencies $\omega_1$ and $\omega_2$
and $N_1=N_2$,
the absorption spectrum 
and the oscillator strength
are shown in Fig.2. Although {\em all} absorption lines are
effected by the dot interaction (represented by the interaction parameter $p$),
and {\em all} modes are optically active,
there is  no {\em qualitative} effect of interaction in the position 
of the absorption lines.
The reason can be understood easily.
In this particular case, the four eigenmodes can be calculated
analytically providing 
\begin{equation}
\omega_{1,2,3,4}=\sqrt{\omega_{eff,i}^2+\left(\frac{\omega_c^*}{2}\right)^2}\pm
\left(\frac{\omega_c^*}{2}\right) \;,\;\;\;(i=1,2)
\label{nonint}
\end{equation}
where
\begin{equation}
\omega^2_{eff,1,2}=\frac{(\omega_1^2+\omega_2^2)}{2}+
\frac{(p_1+p_2)}{2} d \pm 
\sqrt{ \left[ \frac{(\omega_1^2+\omega_2^2)}{2}+\frac{(p_1+p_2)}{2} d  \right]^2 
-(\omega_1^2\; p_2\; d + \omega_2^2\; p_1\; d +\omega_1^2\; \omega_2^2)}
\label{eff}
\end{equation}
(The upper and lower sign belongs to $\omega_{eff,1}$ and $\omega_{eff,2}$, 
respectively).
Consequently, if we had to interpret an experimental spectrum, we
could do this using the formula (\ref{nonint}) 
for non-- interacting dots, but with
the effective (i.e. interaction affected) confinement parameters defined in
(\ref{eff}). 
Only if we take the intensities into account, we see some qualitative effect.
Whereas for non-- interacting dots (with 
$p=0$) and for $B=0$ the oscillator strength of all modes agree
(for a single oscillator, $f$ is independent of
the oscillator frequency),
there is a large difference for interacting dots at $p=0.5$.
This large difference can be understood as follows. In the limit $p\rightarrow
\infty$, the upper pair of modes develops into the spurious Brillouin zone
boundary mode, which has vanishing oscillator strength and the sum rule
has to be fulfilled only by the lower pair (see also the discussion below).\\

In Fig.2 both dot species bare the same number of electrons.
Therefore, only one interaction parameter $p$ is involved.
Calculations with different
$N_i$  (and $p_i$) do not show any qualitative difference.
In the limit of large $p$ (and equal electron numbers) 
we obtain from (\ref{eff})
\begin{equation}
\omega^2_{eff,1,2}=\frac{(\omega_1^2+\omega_2^2)}{2}+
\bigg\{ \begin{array}{c} 2\; pd\\ 0 \end{array}
\pm \frac{(\omega_1^2-\omega_2^2)}{8\; pd} +O(p^{-3})
\label{large-p}
\end{equation}
Consequently, the square of the smaller effective confinement frequency
(which is the only one
giving rise  to modes
with a finite oscillator strength for large $p$)
approaches the mean value of both squared bare confinement frequencies,
whereas the larger one grows continously for large $p$.\\
In Fig.3 and 4b we show the results  for two identical, but mutually rotated,
{\em elliptical dots}. Without dot interaction ($p=0$), we have two doubly
degenerate lines. With increasing interaction strength,
we observe a splitting of degenerate modes
and an anti-crossing behavior for finite $B$.
As in the case of circular dots, the oscillator strength at $B=0$
for non-- interacting dots
($p=0$) agree for all
four modes.
The dot interaction lifts this degeneracy. Additionally,
we observe at $p=0.5$ that the oscillator strength
in the limits of  small and large magnetic fields
is considerable only for two of the modes, except in the gap region,
where three modes contribute.
By comparison of Fig.s 3a and 4b we see that the magnetic field for minimum gap
(between
the second and third mode)
increases with increasing $p$, whereas the gap width decreases.
Consequently, the location and width of the gap provides information on the
interaction strength.\\
By comparison of Fig.s 2 and 3 with Fig.4, and more clearly by consideration of
formula (\ref{large-p}) and Fig.5, it becomes clear that in either case the
lower pair of degenerate modes at $B=0$ converges to a constant
(the mean square bare
confinement frequency $\sqrt{(\omega_1^2+\omega_2^2)/2}$,
which amounts to $1.02 \;\omega_0$ in our numerical example).
Even for finite $B$, there are two branches, which converge to a finite
($B$-dependent) value for $p \rightarrow \infty$, or in other words,
which become
independent of $p$ in this limit.
At first sight this looks surprising because the e e interaction
does not show any saturation, if we increase the interaction parameter,
but it continues to compress
the dot state.
However, there is a simple visual explanation for this feature:
Generally, the
dot interaction adds an additional second order contribution to the
confinement, which has the same
symmetry as the lattice, i.e. it is circular for a cubic lattice. For large
$p$, this additional term outweighs the bare confinement, and the effective
confinement in both dots becomes isotropic and equal.
Thus, we approach the case of a lattice of identical dots, for which
a pair of Kohn modes exists. Because these Kohn modes do not exactly 
agree with the
 modes of noninteracting dots, we call them {\em pseudo} Kohn modes.
In a sense, the Generalized Kohn Theorem reentries for dot lattices
with strong interdot interaction.
In other lattices with lower symmetry, the effective confinement
in the strong interaction limit might
be elliptical, leading pseudo Kohn modes with a gap at $B=0$.
The other pair of modes (which diverge for $p \rightarrow \infty$)
turns into the in-folded modes at the Brillouin zone corner
(because the units cell halves if all dots become equivalent).
These modes become
spurious in the long wavelength and the large--$p$ limit and the oscillator
strength of them converge to zero.\\
In Fig.s 3a and 4b we observe an additional qualitative 
effect of dot interaction.
For isolated elliptical dots we expect a gap between the two
excitation branches at $B=0$. However, for larger $p$ only the pseudo Kohn mode
might be observable, because the oscillator  strength of the BZ boundary mode
decrease rapidly. On the other hand, the two lower modes  for finite $p$
develope out of the degenerate lower mode for $p=0$, whereby the degeneracy
at $B=0$ survives. Therefore, at $B=0$ it looks as if we had a circular dot.
The closing of the gap between the two most intensive branches at $B=0$ is
{\em not} a gradual effect proceeding with
increasing $p$, but initiated by symmetry.
(For a deeper understanding see also the additional  figures in
Ref. \onlinecite{Taut-condmat}.)

\newpage

\begin{figure}[thb]
\vspace{-2cm}
\begin{center}
{\psfig{figure=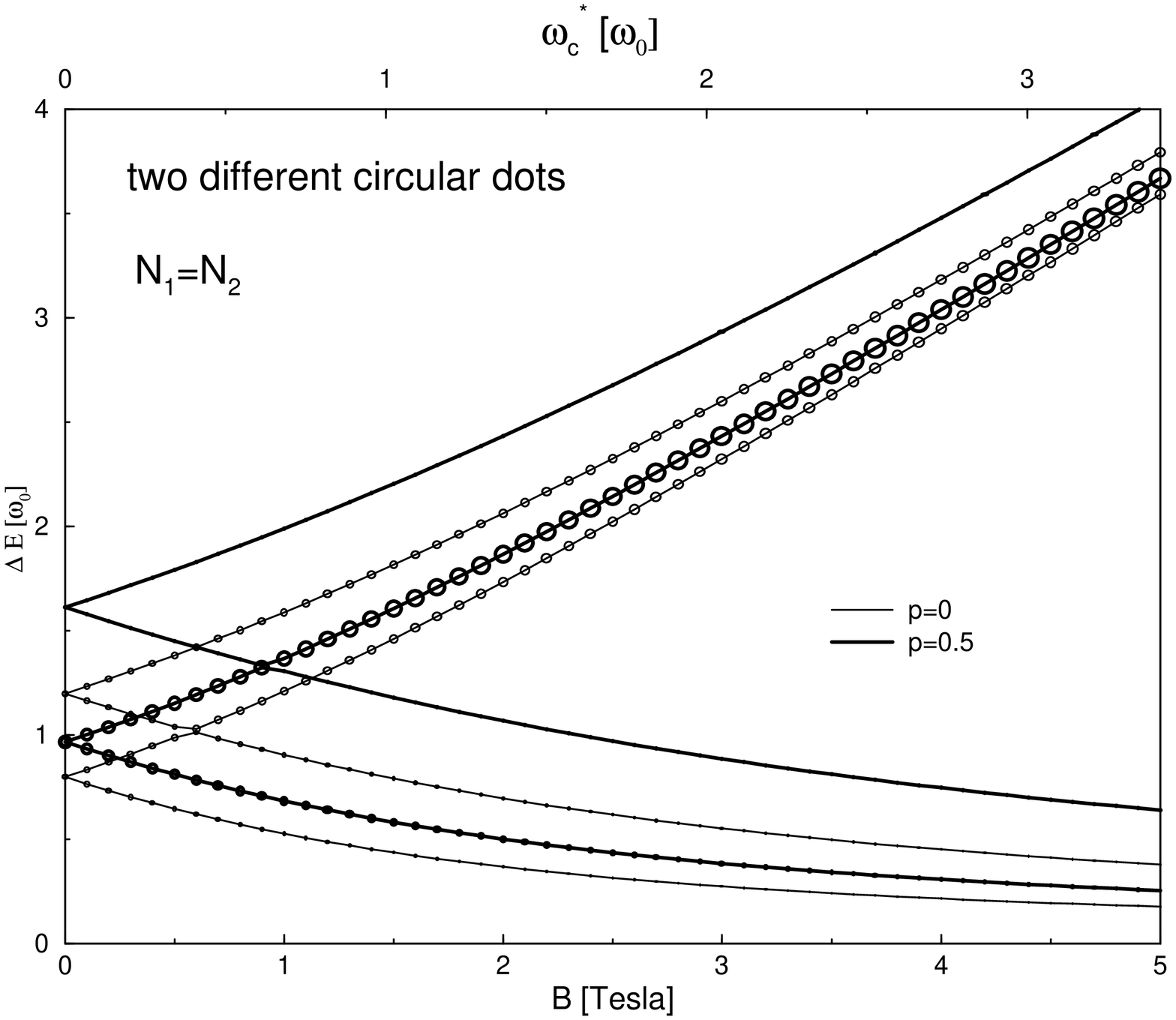,angle=-90,width=9.cm,bbllx=15pt,bblly=45pt,bburx=580pt,bbury=750pt}}
{\psfig{figure=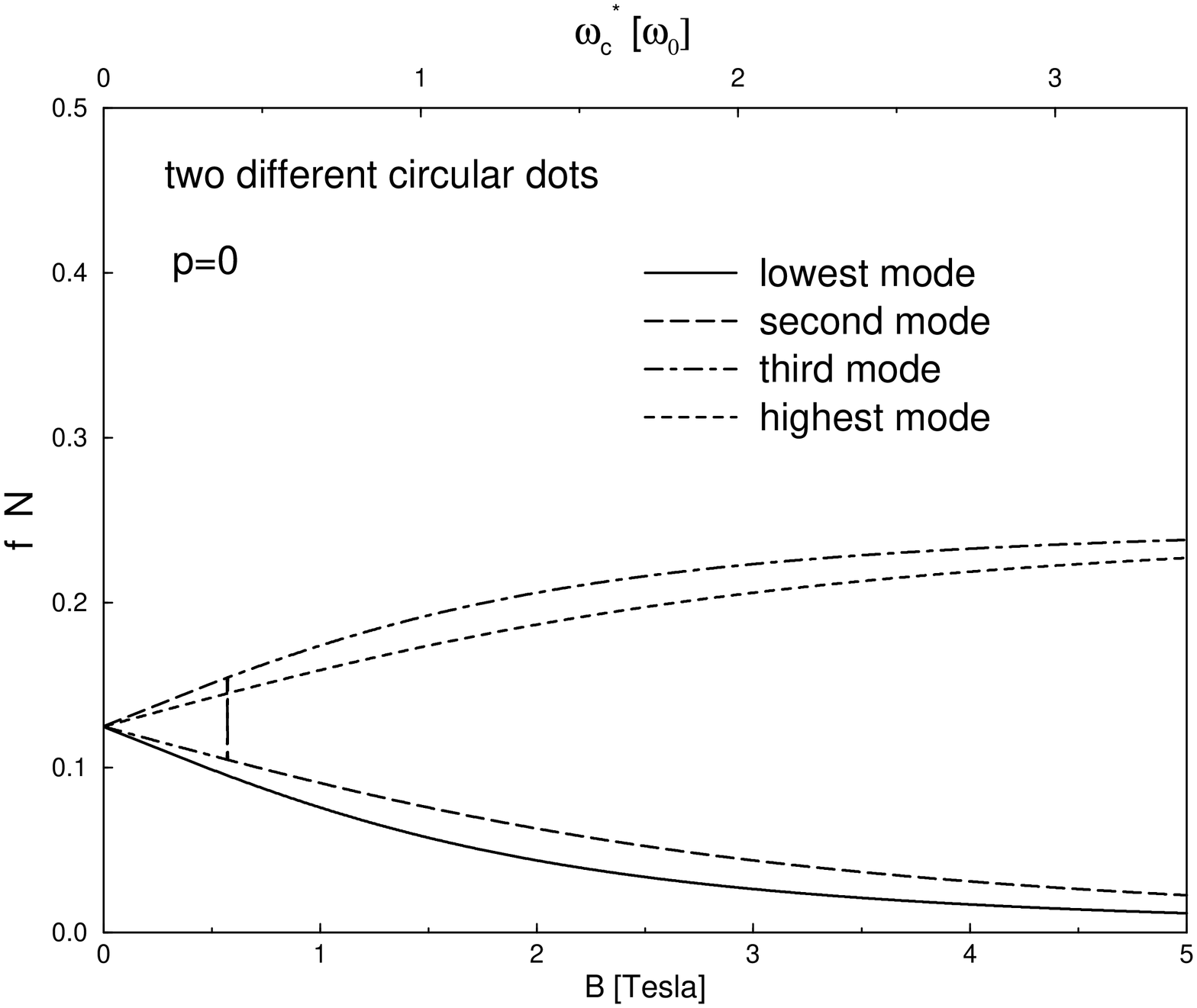,angle=-90,width=9.cm,bbllx=15pt,bblly=45pt,bburx=580pt,bbury=750pt}}
{\psfig{figure=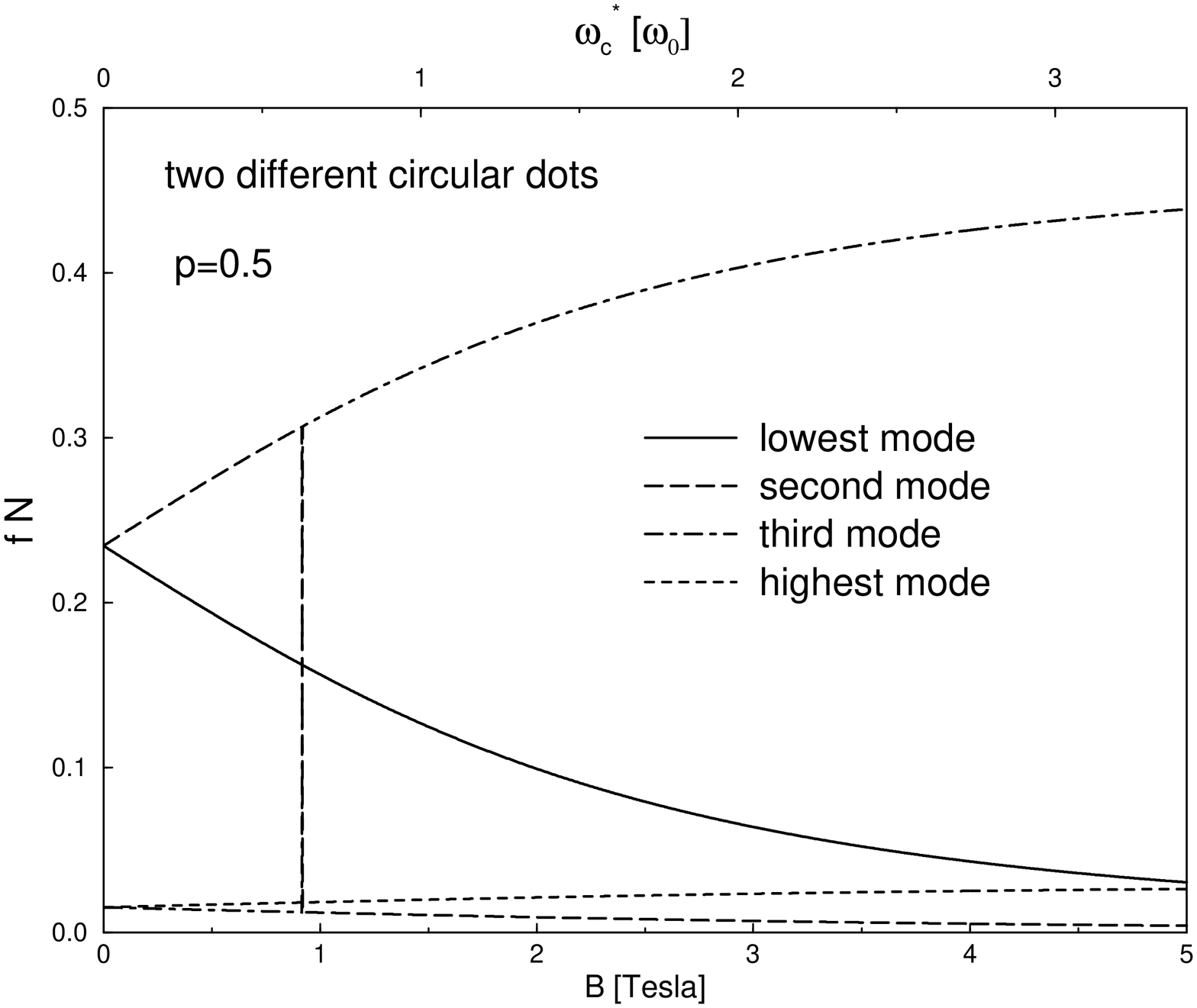,angle=-90,width=9.cm,bbllx=15pt,bblly=45pt,bburx=580pt,bbury=750pt}}
\caption[]{
Excitation modes (a) and oscillator strength (multiplied with $N$)
 for $p=0$ (b)  and  $p=0.5$  (c)
for a lattice with
two different circular dots as described in the text.
The radius of the circles in (a) is proportional to the  oscillator strength
and provides a rough overview.
}
\label{Fig.2}
\end{center}
\end{figure}
\newpage

\begin{figure}[thb]
\vspace{-2cm}
\begin{center}
{\psfig{figure=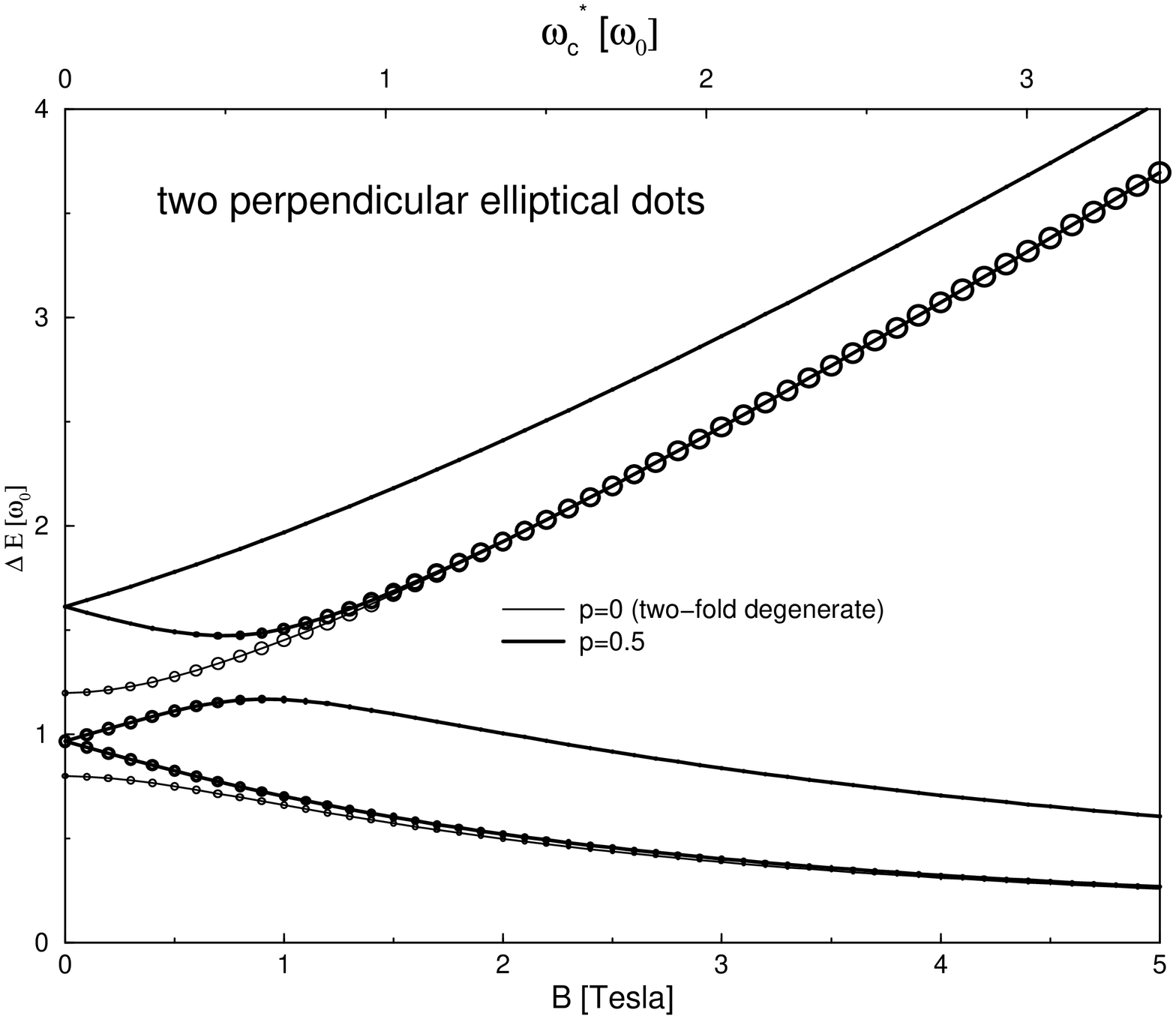,angle=-90,width=9.cm,bbllx=15pt,bblly=45pt,bburx=580pt,bbury=750pt}}
{\psfig{figure=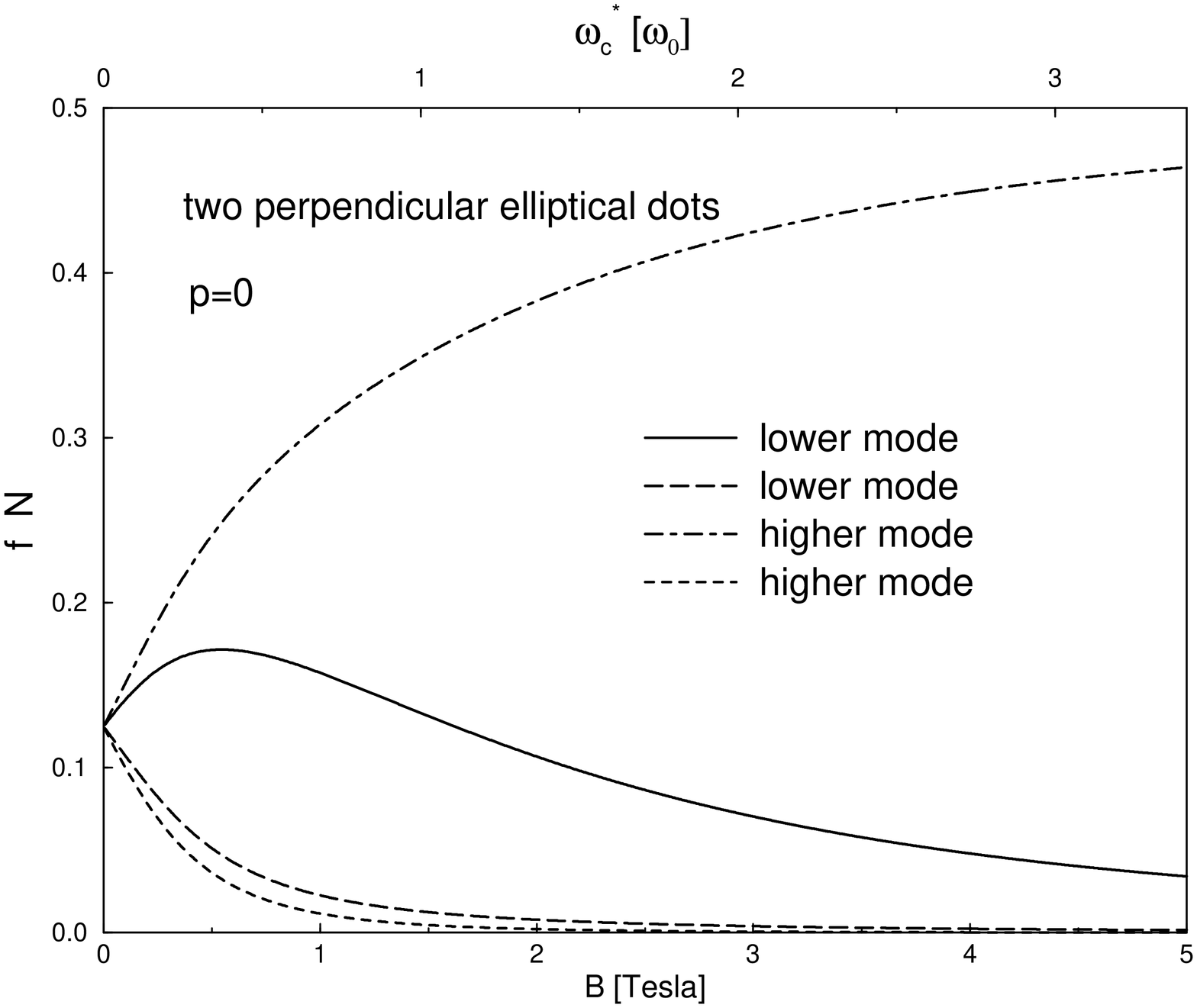,angle=-90,width=9.cm,bbllx=15pt,bblly=45pt,bburx=580pt,bbury=750pt}}
{\psfig{figure=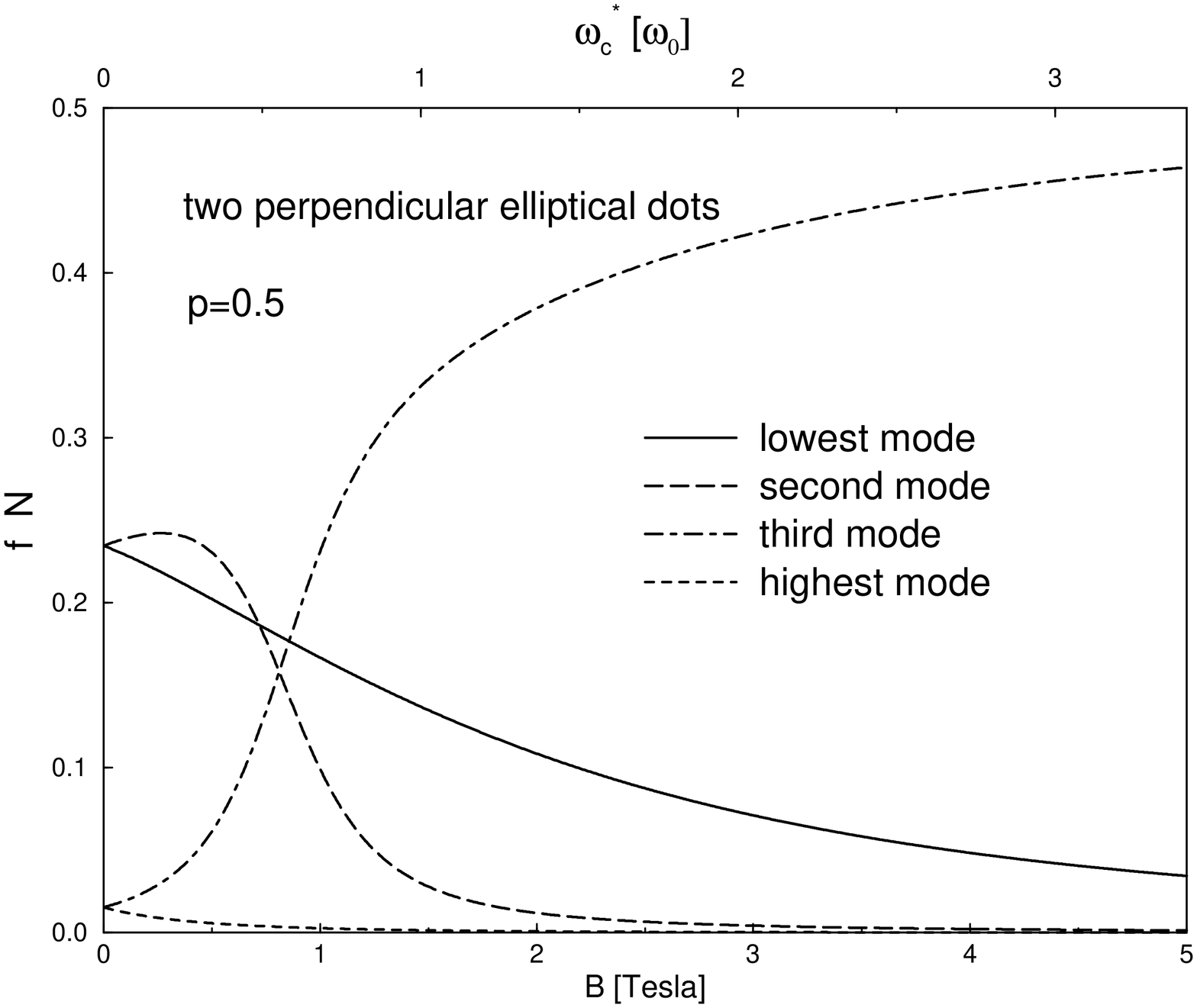,angle=-90,width=9.cm,bbllx=15pt,bblly=45pt,bburx=580pt,bbury=750pt}}
\caption[]{
Excitation modes (a) and oscillator strength (multiplied with $N$)
  for $p=0$ (b) and $p=0.5$ (c)
for a lattice with
two identical, but rotated elliptical dots as described in the text and 
shown in Fig.1.
The radius of the circles in (a) is proportional to the oscillator strength
and provides a rough overview.
}
\label{Fig.3}
\end{center}
\end{figure}
\newpage

\begin{figure}[h]
\vspace{-2cm}
\begin{center}
{\psfig{figure=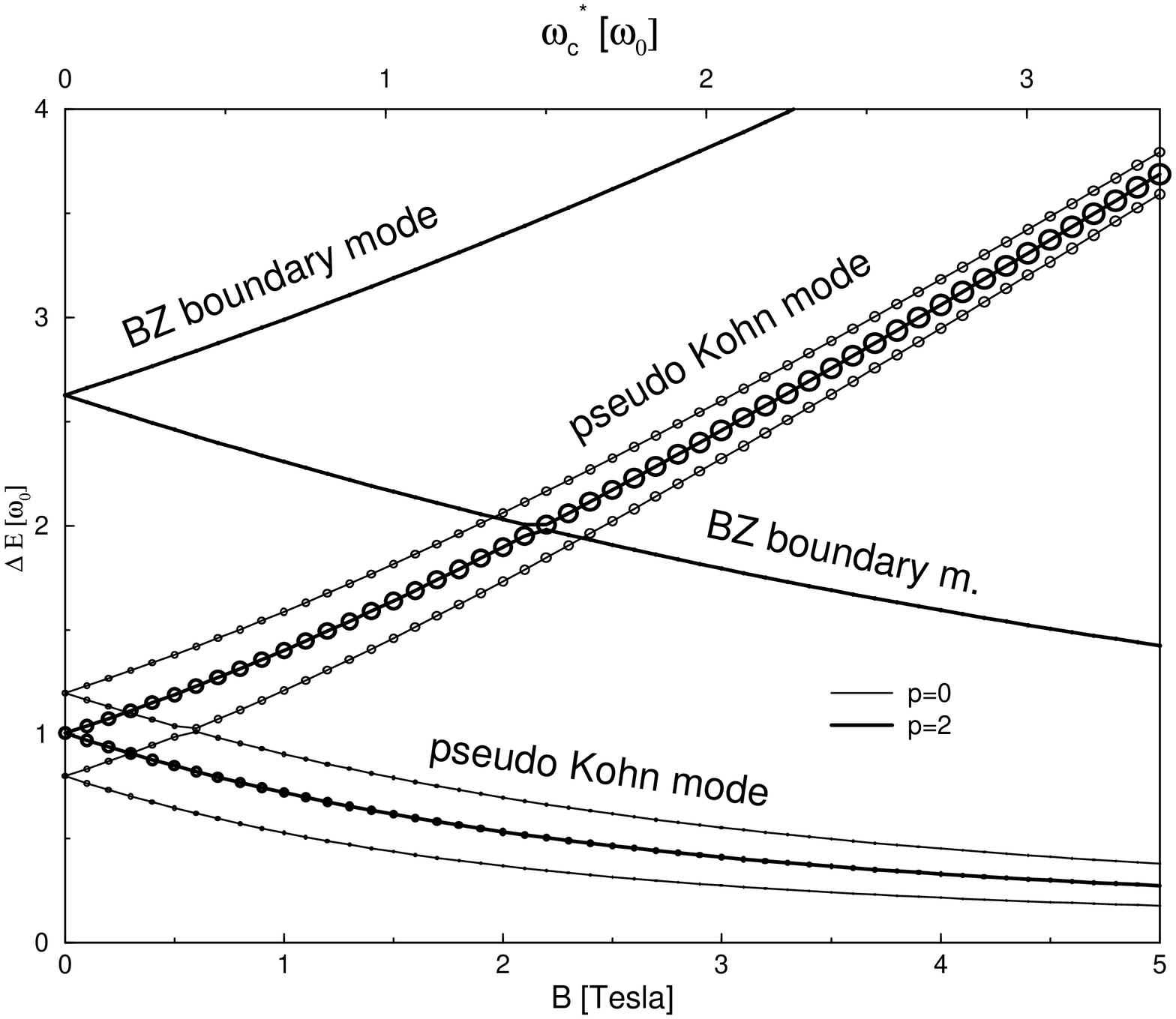,angle=-90,width=12.cm,bbllx=15pt,bblly=45pt,bburx=580pt,bbury=750pt}}
{\psfig{figure=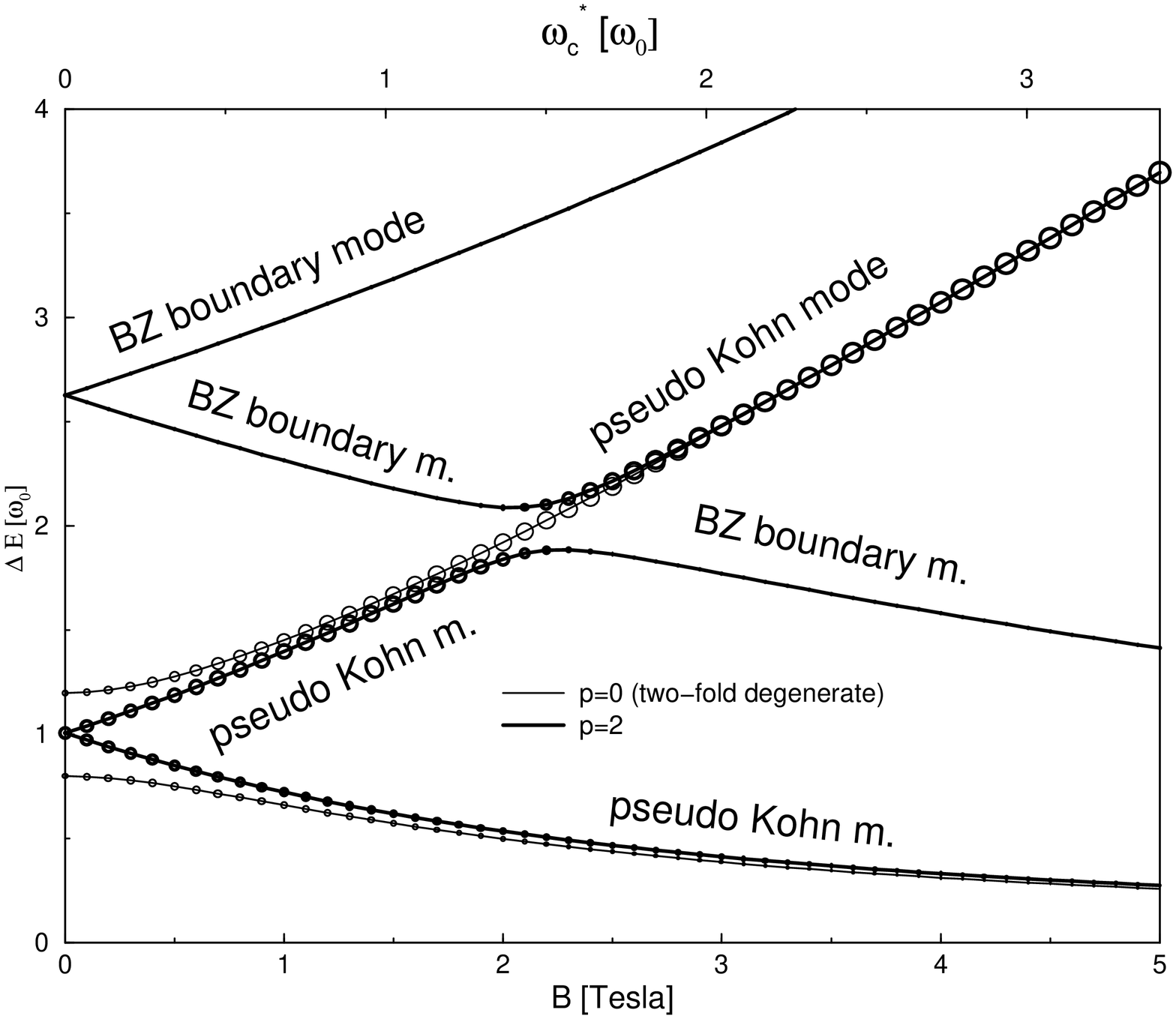,angle=-90,width=12.cm,bbllx=15pt,bblly=45pt,bburx=580pt,bbury=750pt}}
\caption[ ]{
Excitation modes for a lattice with two different circular dots (a) and
two rotated elliptical dots (b) for a large interaction parameter ($p=2$).
The radius of the circles is proportional to the corresponding oscillator strength.
}
\label{Fig.4}
\end{center}
\end{figure}

\begin{figure}[h]
\vspace{-2cm}
\begin{center}
{\psfig{figure=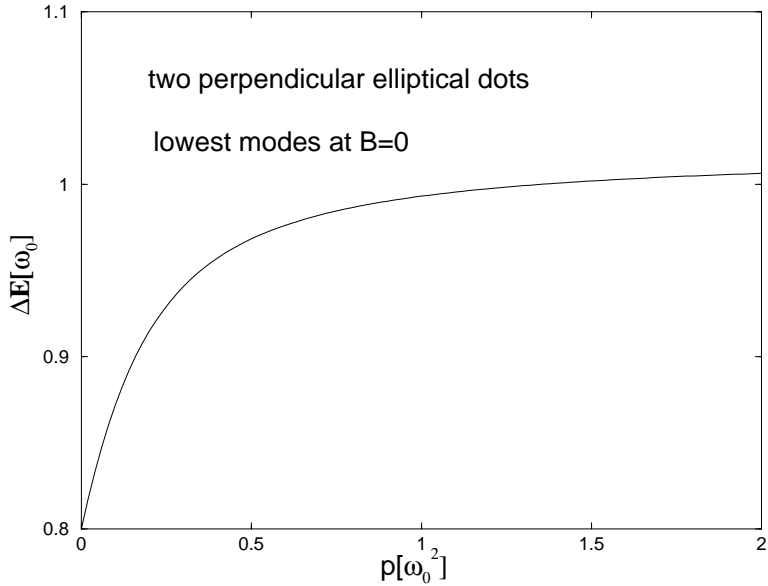,angle=-90,width=12.cm,bbllx=15pt,bblly=45pt,bburx=580pt,bbury=750pt}}
\caption[ ]{
Pseudo Kohn mode at $B=0$ as a function of interaction parameter $p$ 
for a lattice with
two perpendicular elliptical dots per unit cell.
}
\label{Fig.5}
\end{center}
\end{figure}

\section{Summary} 
We have shown that breaking the GKTh by constructing
quantum dot lattices with at least two different dot confinements
 per unit cell has experimentally observable consequences. 
Generally speaking, there are no Kohn modes, i.e. interaction independent modes,
 anymore.
In both of the considered cases, the degeneracy in the FIR {\em intensities}
 at $B=0$ 
between the upper and lower absorption lines is lifted due to dot interaction.
For two mutually rotated elliptical dots (per cell), we observe also a 
splitting of formerly degenerate absorption {\em frequencies} 
and the appearance of an anticrossing.
For two different circular dots no qualitative effect of e e 
interaction in the absorption frequencies is observed. 
Instead, the absorption spectrum can be mimiced by
two noninteracting dots with modified (effective) confinements.
We also pointed out that an extensively strong interaction destroys 
the effect of interaction by producing pseudo-- Kohn modes. Although
this limit cannot be reached experimentally, it might be important to 
take this tendency into consideration.\\
Only in the case of two circular dots there is a simple analytical 
closed form solution.
However,
 with the formulae presented above, the  absorption frequencies and
oscillator strength for any cubic lattice with two
harmonic dot species can be
easily calculated. The only numerical task is to find the eigenvalues
of an explicitly given non--  Hermitian $8\times8$ matrix and to perform a 
special sum
over the eigenvector components.

\section{Acknowledgment}

I am indebted to D.Heitmann, J.Kotthaus, and H.Eschrig,  and their groups,
as well as G.Paasch for helpful discussion.

\end{document}